\documentclass[journal,twoside,web]{ieeecolor}
\usepackage{tmi2}
\usepackage{cite}
\usepackage{amsmath,amssymb,amsfonts}
\usepackage{algorithmic}
\usepackage{graphicx}
\usepackage{float}
\usepackage{adjustbox}
\usepackage{caption}
\usepackage{stfloats}
\usepackage[dvipsnames]{xcolor}
\usepackage{textcomp}
\usepackage[utf8]{inputenc}
\usepackage{longtable}
\usepackage{url}
\usepackage{xr}
\usepackage{multirow}
\usepackage[linesnumbered,ruled,vlined]{algorithm2e}
\def\baselinestretch{1.0}
\SetCommentSty{mycommfont}

\SetKwInput{KwInput}{Input}                
\SetKwInput{KwOutput}{Output}              
\usepackage[font=small,justification=justified,belowskip=2pt,aboveskip=2pt]{caption}
\definecolor{bondiblue}{rgb}{0.0, 0.58, 0.71}
\definecolor{brightcerulean}{rgb}{0.11, 0.62, 0.74}

\usepackage[math]{cellspace}
\floatname{algorithm}{Algorithm}

\def\SB#1{\textsubscript{#1}}
\def\BibTeX{{\rm B\kern-.05em{\sc i\kern-.025em b}\kern-.08em
    T\kern-.1667em\lower.7ex\hbox{E}\kern-.125emX}}
\begin{document}
\title{Federated Learning of Generative Image Priors for MRI Reconstruction\vspace{-0.25cm}}
\author{Gokberk Elmas, Salman UH Dar, Yilmaz Korkmaz, Emir Ceyani, Burak Susam, Muzaffer Ozbey, \\  Salman Avestimehr, and Tolga \c{C}ukur \vspace{-1.25cm}
\\
\thanks{\\
This study was supported in part by a TUBA GEBIP 2015 fellowship, and a BAGEP 2017 fellowship (Corresponding author: Tolga Çukur).}
\thanks{G. Elmas, S. UH. Dar, Y. Korkmaz, M. Ozbey, B. Susam and T. Çukur are with the Department of Electrical and Electronics Engineering, and the National Magnetic Resonance Research Center, Bilkent University, Ankara, Turkey (e-mail: \{gokberk@ee, salman@ee, korkmaz@ee, muzaffer@ee, burak.susam@ug, cukur@ee\}.bilkent.edu.tr).}
\thanks{E. Ceyani and S. Avestimehr are with the Department of Electrical and Computer Engineering, University of Southern California, Los Angeles, CA 90089 USA (e-mail: \{ceyani@, avestime@\}usc.edu).}
}

\maketitle

\begin{abstract}
Multi-institutional efforts can facilitate training of deep MRI reconstruction models, albeit privacy risks arise during cross-site sharing of imaging data. Federated learning (FL) has recently been introduced to address privacy concerns by enabling distributed training without transfer of imaging data. Existing FL methods employ conditional reconstruction models to map from undersampled to fully-sampled acquisitions via explicit knowledge of the accelerated imaging operator. Since conditional models generalize poorly across different acceleration rates or sampling densities, imaging operators must be fixed between training and testing, and they are typically matched across sites. To improve generalization and flexibility in multi-institutional collaborations, here we introduce a novel method for MRI reconstruction based on Federated learning of Generative IMage Priors (FedGIMP). FedGIMP leverages a two-stage approach: cross-site learning of a generative MRI prior, and subject-specific injection of the imaging operator. The global MRI prior is learned via an unconditional adversarial model that synthesizes high-quality MR images based on latent variables. A novel mapper subnetwork produces site-specific latents to maintain specificity in the prior. During inference, the prior is first combined with subject-specific imaging operators to enable reconstruction, and it is then adapted to individual cross-sections by minimizing a data-consistency loss. Comprehensive experiments on multi-institutional datasets clearly demonstrate enhanced generalization performance of FedGIMP against both centralized and federated methods based on conditional models.
\vspace{-0.2cm}
\end{abstract}

\begin{IEEEkeywords}
MRI, accelerated, reconstruction, generative, prior, federated learning, distributed, collaborative. \vspace{-0.3cm}
\end{IEEEkeywords}

\bstctlcite{IEEEexample:BSTcontrol}

\section{Introduction}
\IEEEPARstart{M}{agnetic} Resonance Imaging (MRI) is a principal radiological modality owing to its non-invasiveness and exceptional soft-tissue contrast. Yet, an inevitable consequence of its low signal-to-noise ratio (SNR) efficiency is prolonged exams that hinder clinical use. Accelerated MRI methods based on undersampled acquisitions improve efficiency by recovering missing data via reconstruction algorithms that incorporate additional prior information \cite{Lustig2007,Zhao2015}. Deep learning models have been adopted for MRI reconstruction, given their strong ability to capture data-driven priors for inverse problems \cite{Wang2016,Hammernik2017,Kwon2017,Dar2017,Han2018a,ADMM-CSNET,raki,Xiang2019}. Deep reconstruction models are typically trained to perform a conditional mapping from undersampled acquisitions to images that are consistent with respective fully-sampled acquisitions \cite{Schlemper2017,MoDl,Quan2018c,KikiNet,Mardani2019b,Polakjointvvn2020,rgan,FengNNLS2021,PatelRecon}. As such, successful learning procedures involve training on a large collection of input-output data \cite{LiangSPM}. Unfortunately, economic and labor costs along with patient privacy concerns can prohibit compilation of diverse datasets centralized at a single institution \cite{Kaissis2020}.

Aiming at this limitation, federated learning (FL) is a promising framework that facilitates multi-institutional collaborations via decentralized training of learning-based models \cite{WenqiLi2019,Sheller2019,Rieke2020,Roth2020,Li2020,Liu2021}. An FL server periodically collects locally-trained model parameters from individual sites in order to obtain a shared global model across sites \cite{McMahan2017CommunicationEfficientLO,fl_opt_guide}. Following aggregation of local models, the global model is then broadcast back onto individual sites for continual training. This decentralized procedure allows the model to be trained on multiple datasets without sharing of local data, thereby mitigating privacy concerns \cite{Li2020FederatedLC}. Unfortunately, multi-institutional datasets can show substantial heterogeneity in the MR image distribution (e.g., due to different tissue contrasts, scanners) or the accelerated imaging operator (e.g., due to different acceleration rates, sampling densities) \cite{fastmri}. In turn, FL-trained models can suffer from performance losses under significant domain shifts across sites, or across the training and test sets \cite{KnollGeneralization,Liu2021}. 

Few recent studies on FL-based MRI reconstruction have considered domain shifts across sites. In \cite{guo2021}, adversarial alignment between source and target sites was proposed to improve similarity of latent-space representations in reconstruction models. In \cite{feng2021}, reconstruction models were split into a global encoder shared across sites, followed by local decoders trained separately at each site. Despite improved performance against across-site variability, both methods are based on conditional reconstruction models that assume explicit knowledge of the imaging operator. Conditional models generalize poorly under changes to the imaging operator \cite{Zhu2018,Biswas2019}. In turn, matching acceleration rates and sampling densities must be prescribed between the training and test sets, and similar prescriptions are often utilized among sites \cite{guo2021}. This requirement can restrict flexibility in multi-institutional collaborations and necessitate model retraining under significant changes to the desired imaging operator \cite{KnollGeneralization}. 

\begin{figure*}[t]
\vspace{-0.45cm}
\centering
\includegraphics[width=0.900\textwidth]{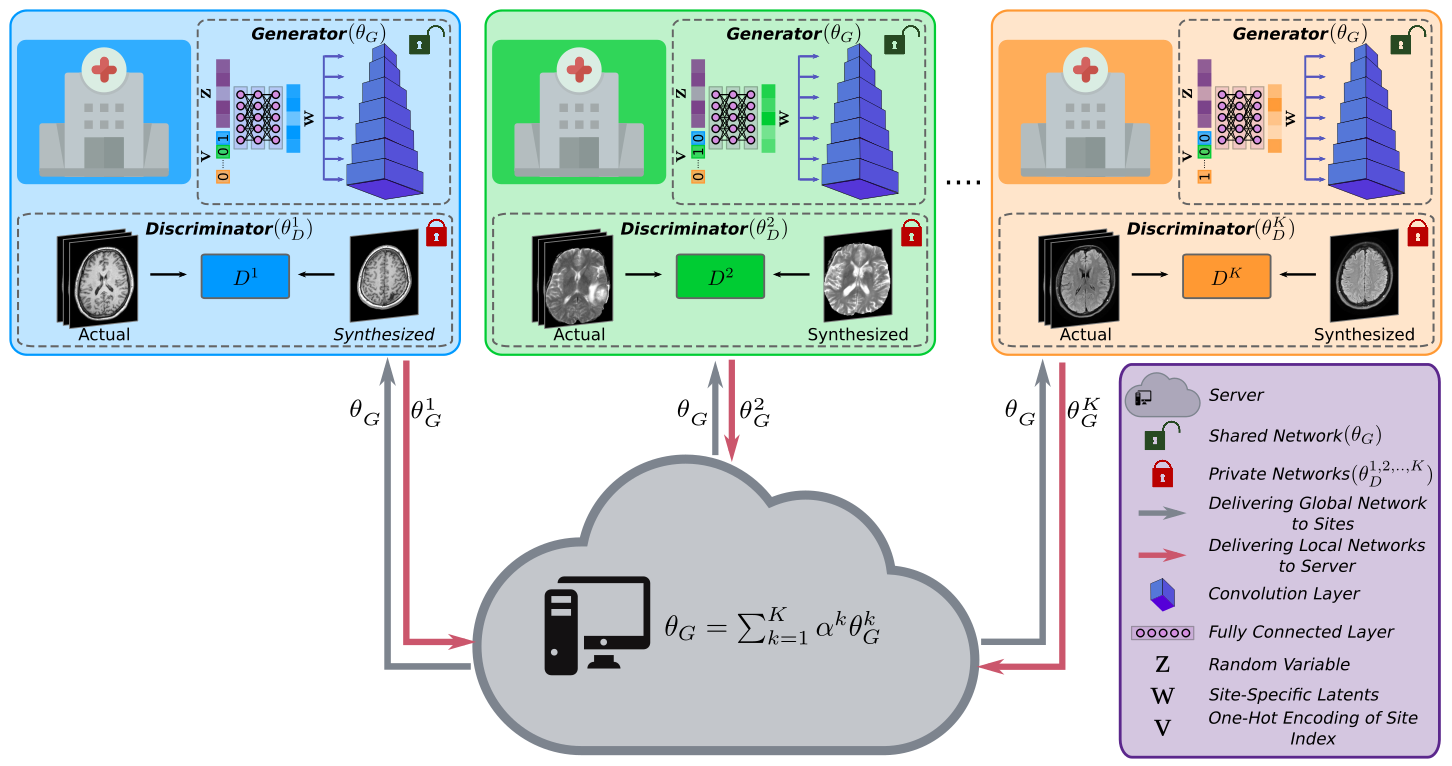}
\captionsetup{justification=justified, width=\linewidth}
\caption{FedGIMP is a decentralized reconstruction method based on federated learning of a generative MRI prior decoupled from the imaging operator. The prior is embodied as an unconditional adversarial model that synthesizes high-quality MR images given site-specific latent variables. Cross-site-shared generators of parameters $\theta_G$ are used along with site-specific discriminators of parameters $\theta_{D}^{1,..,K}$. In each communication round, sites perform local updates to $\theta_G$ to minimize a synthesis loss, and the server then aggregates updated models.
\vspace{-0.1cm}}
\label{fig:fedgimp_main}
\end{figure*}

Here, we introduce a novel method for accelerated MRI reconstruction based on Federated learning of Generative IMage Priors (FedGIMP). Unlike previous methods, we propose a two-stage reconstruction approach: cross-site learning of a generative MRI prior (Fig. \ref{fig:fedgimp_main}), and subject-specific injection of the imaging operator (Fig. \ref{fig:fedgimp_opt}). The global MRI prior is captured using an unconditional adversarial model that synthesizes high-quality MR images based on low-dimensional latent variables produced by a mapper. The mapper calculates site-specific latent variables to preserve specificity in the prior. In the inference stage, the global MRI prior is combined with subject-specific imaging operators that can vary flexibly across sites and between the training and test sets. By minimizing data-consistency loss on acquired k-space samples, the generative prior is adapted to perform reconstruction of individual test cross-sections. Code for FedGIMP can be found at: {\small \url{https://github.com/icon-lab/FedGIMP}}. 


\vspace{0.5mm}
\subsubsection*{\textbf{Contributions}}
\begin{itemize}
    \item We introduce a novel FL-based MRI reconstruction that learns a generative MRI prior detached from the imaging operator. This approach increases flexibility in multi-site collaborations and improves reliability against domain shifts in the imaging operator.
    
    \item We propose an unconditional adversarial model with site-specific latents and cross-site-shared generator weights to capture both site-specific and site-general representations of high-quality MR images. 
    \item We propose a subject-specific adaptation of the MRI prior to enhance reliability against domain shifts in the MR image distribution between training and test sets.
\end{itemize}

\vspace{-0.25cm}
\section{Related Work}
\vspace{-0.1cm}
Deep MRI reconstruction is pervasively based on conditional models that directly map undersampled acquisitions to images consistent with fully-sampled acquisitions \cite{Hammernik2017}. These models are trained on large sets of paired input-output data under a specific accelerated imaging operator. Heavy data demand limits applicability since curating large datasets at a single site is challenging \cite{data_diff,GuoTMI2021}. To facilitate curation, unpaired \cite{Quan2018c,oh2020,lei2020}, self-supervised \cite{yaman2020,Huang2019self,Tamir2019,FengLiu2021}, or transfer \cite{Dar2017,KnollGeneralization} learning strategies were proposed. However, these methods require centralized training following cross-site data transfer that raises patient privacy concerns \cite{Kaissis2020}. 

Federated learning (FL) is a decentralized framework for multi-institutional collaborations that communicates model parameters instead of raw data \cite{Li2020FederatedLC}. FL distributes costs related to the formation and processing of datasets across sites, while mitigating concerns regarding the data privacy \cite{Kaissis2020}. FL methods have readily been demonstrated on imaging tasks such as segmentation and classification \cite{Sheller2019,WenqiLi2019,Roth2020}. A major consideration is the reliability against domain shifts in multi-site imaging data, collected with varying imaging protocols and scanners. Data harmonization was proposed to remove site-specific variations while emphasizing shared variability across sites \cite{karayumak2019,dewey2019,li2021fedbn}. Although harmonization can improve population-level analysis, it can discard patient-level information of diagnostic value. Episodic learning in frequency space was proposed to improve generalization to unseen test domains for segmentation \cite{Liu2021}. Adversarial alignment and network splitting methods were also proposed for classification \cite{Li2020,VAFL,park2021}. While promising results were reported, it is nontrivial to directly adopt image analysis models for MRI reconstruction that requires image formation from raw data.

Domain shifts in MRI reconstruction involve heterogeneity in the MR image distribution and in the imaging operator, which can elicit performance losses when heterogeneity is prominent across sites, or across the training-test sets \cite{KnollGeneralization}. Few recent FL studies on single-coil MRI reconstruction have considered domain shifts across sites. In \cite{guo2021}, cross-site-shared latent-space representations were obtained by adversarially aligning all sites to the targeted test site in each communication round. In \cite{feng2021}, a split reconstruction model with a global encoder and unshared decoder was used to maintain site-specific and site-general representations. While demonstrating remarkable results, these recent methods are based on conditional models that are explicitly informed on the imaging operator. This can limit reconstruction performance and necessitate model retraining under notable domain shifts in the imaging operator \cite{Zhu2018,Biswas2019}. As such, previous studies have prescribed matching acceleration rates and sampling densities between the training-test sets, and usually across sites. 

Here, we propose federated learning of a generative MRI prior, and reconstruction via prior adaptation after combination with the subject-specific imaging operator. Separation of the image prior from the imaging operator has recently been considered for centralized reconstruction models trained on single-site data \cite{Knoll2019inverseGANs,Konukoglu2019,korkmaz2021unsupervised}, yet it has not been studied in the context of FL. To our knowledge, FedGIMP is the first FL method that decouples the MRI prior from the imaging operator, and the first FL method that reconstructs multi-coil MRI data. These unique aspects enable FedGIMP to effectively address domain shifts due to mismatched imaging operators across separate sites, and across the training-test sets. 

\vspace{-0.2cm}
\section{Theory}
\vspace{-0.1cm}
\subsection{Federated Learning of Conditional MRI Models}
\vspace{-0.1cm}
Accelerated MRI entails reconstruction of an underlying MR image $m$ from undersampled k-space acquisitions $y$:
\begin{equation}
\label{eq:sampling}
A m = y,
\end{equation}
where $A$ is the imaging operator that includes the effects of coil sensitivities and partial Fourier transformation on acquired k-space. As Eq. \ref{eq:sampling} is underdetermined, additional prior information is incorporated to regularize the reconstruction:
\begin{equation}
\label{eq:regularized_recon}
\widehat{m}=\underset{m}{\operatorname{argmin}}\|y-A m \|_{2}^{2} + H(m),
\end{equation}
where $H(m)$ enforces the prior \cite{Lustig2007}. Deep models have recently become the predominant method for solving Eq. \ref{eq:regularized_recon} by training of data-driven priors on diverse datasets \cite{Hammernik2017}. 

In FL, training is performed via communication between multiple sites and a server \cite{Li2020FederatedLC}. The server retains a global model ($C_{\theta}$ with parameters $\theta$), whereas each site retains a local model of matching architecture ($C_{\theta}^k$ for site $k$, where $k \in \{1,..,K\}$). In each communication round, local models are initialized with the global model broadcast from the server, $\theta_C^k \gets \theta_C$, and updated to minimize a local reconstruction loss:
\begin{equation}
\label{eq:fed_dnn_training}
\mathcal{L}^{k}_{Rec}(\mathcal{D}^k,A^k_{tr};\theta^k)= \mathbb{E}_{(m^{k}_{tr},y^{k}_{tr})} \left[\|m^{k}_{tr} - C_{\theta^k}(A^{\dagger k}_{tr}y^{k}_{tr}) \|_{2} \right],
\end{equation}
where $\mathbb{E}$ denotes expectation, $\mathcal{D}^k$ are local training data comprising undersampled acquisitions ($y^k_{tr}$) and reference images obtained from fully-sampled acquisitions ($m^k_{tr}$), and $A^k_{tr}$, $A^{\dagger k}_{tr}$ are the imaging operator and its adjoint. $C$ is a conditional model with parameters $\theta^k$ that receives zero-filled Fourier reconstruction of $y^k_{tr}$. At the end of each round, updated local models are aggregated via federated averaging (FedAvg) \cite{McMahan2017CommunicationEfficientLO}:
\begin{equation}
\label{eq:fed_avg}
\theta =  \sum_{k=1}^{K} \alpha^k \theta^k,
\end{equation}
where $\alpha^k$ denotes the relative site weights typically set to $\frac{N^k}{N}$, where $N$ is the total number of training samples across all sites and $N^k$ is the number of training samples at the $k^{th}$ site. 

The trained global model ($C_{\theta^*}$) is then used for inference:
\begin{equation}
\label{eq:fed_ref}
\hat{m}_{k,s} =  C_{\theta^*}(A_{test}^{\dagger k}y^{k,s}_{test}),
\end{equation}
where $\hat{m}_{k,s}$ is the reconstruction and $y^{k,s}_{test}$ is the undersampled acquisition for the $s^{th}$ subject at the $k^{th}$ site, and $A_{test}^{\dagger k}$ is the adjoint imaging operator at site $k$. Since conditional models generalize poorly against heterogeneity in the imaging operator, $A_{tr}^k$ and $A_{test}^k$ are typically matched across training-test sets (i.e. across training and test subjects), and across sites.

\begin{figure}[t]
\centering
\includegraphics[width=0.85\columnwidth]{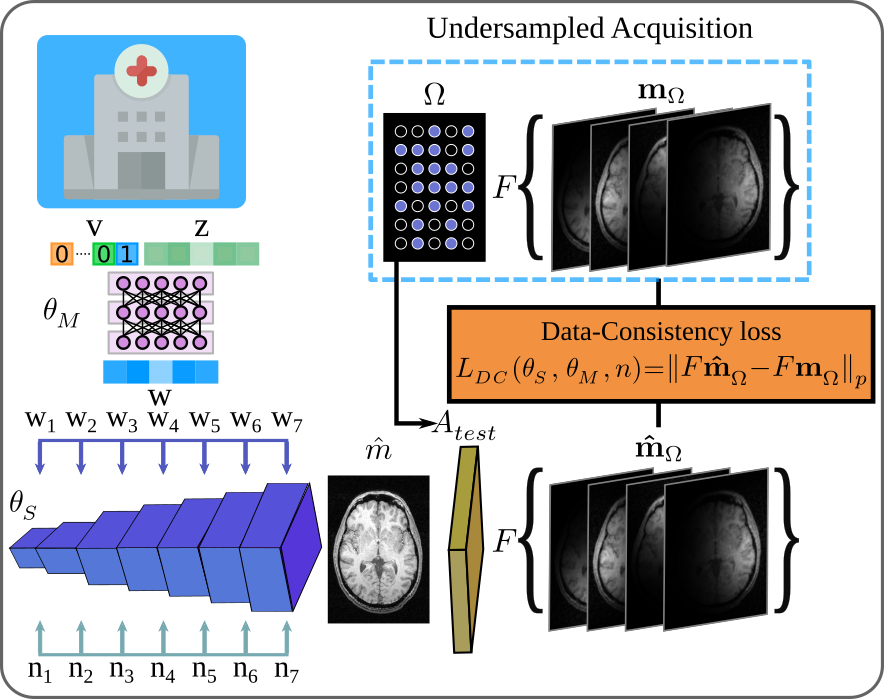}
\captionsetup{justification=justified, width=\columnwidth}
\caption{FedGIMP's MRI prior is embodied as a generator that synthesizes a high-quality, coil-combined MR image ($\hat{m}$). To perform reconstruction, the trained generator is combined with the subject-specific imaging operator at a test site ($A_{test}$) and adapted to minimize a data-consistency loss on acquired k-space samples ($L_{DC}$). $L_{DC}$ is expressed by projecting $\hat{m}$ onto individual coils, undersampling multi-coil images in k-space (with the pattern $\Omega$), and comparing synthesized and acquired k-space samples. For each cross-section, inference optimization is conducted over the synthesizer and mapper parameters ($\theta_{S,M}$) and noise ($n$).}
\label{fig:fedgimp_opt}
\vspace{-0.1cm}
\end{figure}

\vspace{-3mm}
\subsection{Federated Learning of Generative MRI Priors}
\vspace{-1mm}
To improve flexibility in multi-site collaborations, we propose a novel method based on decentralized learning of generative MRI priors. A global MRI prior is trained using an unconditional adversarial model with shared generator and local discriminator networks (Fig. \ref{fig:fedgimp_main}). During inference, the prior is combined with subject-specific imaging operators and adapted to the reconstruction task (Fig. \ref{fig:fedgimp_opt}). The proposed architecture and learning procedures are described below.

\vspace{1mm}
\subsubsection{Unconditional Adversarial Model}
FedGIMP employs an unconditional architecture for generative modeling of high-quality, coil-combined MR images. Here, we propose a style-generative model as inspired by the success of this model family in natural image synthesis \cite{StyleGAN1}. Our proposed model has a generator to synthesize realistic MR images, and a discriminator to differentiate between synthetic and actual MR images. To preserve specificity, the generator comprises a novel mapper to compute site-specific latent variables, which are transformed onto images via a synthesizer subnetwork.

\begin{algorithm}[t]
\small
\caption{FedGIMP training}\label{alg:flgimp_training}

\KwInput { $\mathcal{D}=\{\mathcal{D}^1,...,\mathcal{D}^K\}$: datasets from $K$ sites. \\
$L$: number of communication rounds.\\ 
$I$: number of local epochs. \\
\{$G^1,...,G^K$\}: local generators of params. \{$\theta_G^1,...,\theta_G^K$\}. \\
\{$D^1,...,D^K$\}: local discriminators of params. \{$\theta_D^1,...,\theta_D^K$\}. \\
\{$\alpha^1,...,\alpha^K$\}: weighting coefficients of K sites. \\
$G$: global generator of params. $\theta_G$. \\
$Opt()$: optimizer that computes parameter updates. \\
}

\KwOutput{$G$ trained global model of params. $\theta_{G}^*$.} 
 Initialize parameters\\
\For{$l=1:L$}
{
    \For{$k=1:K$}
    {
        $\theta_G^k \gets \theta_G$ (receive global generator)\\
        \For{$i=1:I$}
        {
            Calculate $\nabla_{\theta_G^k}\mathcal{L}^k_{G}(\theta_G^k)$ based on Eq. \ref{eq:fedgimp_synthesizer_loss}\\
            $\theta^{k}_{G} \gets \theta^{k}_{G}- Opt(\nabla_{\theta_G^k}\mathcal{L}^k_{G})$ \\
            Calculate $\nabla_{\theta_D^k}\mathcal{L}^k_{D}(\mathcal{D}^k_i;\theta_D^k)$ based on Eq. \ref{eq:fedgimp_discriminator_loss} \\
            $\theta^{k}_{D} \gets \theta^{k}_{D}- Opt(\nabla_{\theta_D^k}\mathcal{L}^k_{D})$ \\
        }
    }
    $\theta_G \gets  \sum_{k=1}^{K} \alpha^k \theta_G^k$ (aggregate local generators)\\
}
\Return $\theta^*_{G}=\theta_{G}$ \\
\end{algorithm}

\textbf{Generator ($G$)}: The generator uses mapper ($M$) and synthesizer ($S$) subnetworks to transform low-dimensional random variables onto high-dimensional MR images. \underline{$M$} receives samples from i.i.d. normal variables $z \in \mathbb{R}^{1\times J}$ augmented with site index as a one-hot encoding vector $v\in \mathbb{R}^{1\times K}$. It computes latent variables $w\in \mathbb{R}^{1\times J}$:
\begin{align}
w=M_{\theta_M}(z \oplus v),
\end{align}
where $\oplus$ denotes concatenation. At the $i^{th}$ layer of $M$ with $L_M$ layers, latent variables $z_i$ are mapped onto $z_{i+1}$ as:
\begin{align}
z_{i+1}=FC_{M,i}(z_{i}),
\end{align}
where $FC_{M,i}$ denotes a fully-connected layer with parameters $\theta_{M,i} \in \mathbb{R}^{(J+K)\times J} \mbox{ if }  i=1;\mathbb{R}^{J\times J} \mbox{ if } i\neq1$. Receiving intermediate latents from $M$, \underline{$S$} generates an MR image progressively across $L_S$ layers. In the $i^{th}$ layer of $S$, feature maps from the preceding layer ($X_{i}^0\in \mathbb{R}^{\frac{h_1}{2}\times \frac{h_2}{2} \times q}$) are two-fold upsampled. The upsampled maps ($X_{i}^1\in \mathbb{R}^{h_1\times h_2\times q}$) are then processed through a serial cascade of blocks: $Conv_S^1$ (convolution), $NI^1$ (noise injection), $AdaIN^1$ (adaptive instance normalization), $Conv_S^2$, $NI^2$, $AdaIN^2$. The first convolution block processes local features, $X_{i}^{2}=Conv_S^1(X_{i}^{1})$:
\begin{align}
X_{i}^{2} =\begin{bmatrix}
    \sum_{c} X_{i}^{1,c} \circledast \theta_{S,i}^{c,1} \\
     \vdots \\
     \sum_{c}  X_{i}^{1,c} \circledast \theta_{S,i}^{c,u} \\
\end{bmatrix},
\label{eq:convolution}
\end{align}
where $\theta_{S,i}^1\in \mathbb{R}^{j\times j \times q \times u}$ are convolution kernels, $u$ is the number of output feature channels, $c$ is the channel index, and $\circledast$ denotes convolution. In Eq. \ref{eq:convolution}, $X_{i}^{2}$ is a 3D tensor and matrix rows span the $3^{rd}$ dimension of $X_{i}^{2}$. Next, the noise-injection block adds scaled noise variables onto feature maps to control low-level structural details, $X_{i}^{3}=NI^1(X_{i}^{2})$:
\begin{align}
X_i^{3} =\begin{bmatrix}
    \varphi(X_i^{2,1} +\epsilon_i^{1,1} n_i^{1}\\
     \vdots \\
     \varphi(X_i^{2,u} +\epsilon_i^{1,u} n_i^{1})\\
\end{bmatrix},
\end{align}
where $n_i^{1} \in \mathbb{R}^{h_1\times h_2}$ is multiplied with the scalar $\epsilon_i^{1,c}$ and added onto the $c^{th}$ channel of $X_i^{2} \in \mathbb{R}^{h_1\times h_2\times u}$, $\varphi$ is an activation function. To control high-level style features, adaptive instance normalization modulates feature maps given intermediate latent variables, $X_{i}^{4}=AdaIN^1(X_{i}^{3},w)$ \cite{adain}:
\begin{align}
X_i^{4} = \gamma^1_i (w) \frac{(X_i^{3}-\mu (X_i^{3}))}{\sigma (X_i^{3})} + \beta^1_i (w),
\end{align}
where $\gamma^1_i \in \mathbb{R}^{u\times 1}$ and $\beta^1_i \in \mathbb{R}^{u\times 1}$ are learnable affine transformations of $w$ that control the scale and bias of each feature channel respectively. $\mu$ and $\sigma$ are mean and variance along the channel dimension of $X_i^{3}$. Note that the cascade of convolution, noise injection and adaptive instance normalization is repeated a second time to yield the mappings $X_{i}^{5}=Conv_S^2(X_{i}^{4})$, $X_{i}^{6}=NI^2(X_{i}^{5})$, $X_{i+1}^{0}=AdaIN^2(X_{i}^{6},w)$. The parameters of these repeat blocks are distinct from those in the initial blocks, except for tied intermediate latents input to $AdaIN^{1,2}$. Overall, the mapping through the generator is: 
\begin{align}
\hat{x}=G_{\theta_G}(z \oplus v,n) = S_{\theta_S}(M_{\theta_M}(z \oplus v),n),
\end{align}
where $\hat{x}$ is the synthesized image, the generator parameters $\theta_G$=\{$\theta_M, \theta_S$\} include mapper and synthesizer parameters.

\textbf{Discriminator ($D$)}: The discriminator receives either an actual or synthetic MR image as input, and attempts to discriminate between the two cases. $D$ contains $L_D-1$ layers cascading two-fold downsampling and convolution blocks ($Conv_D$), followed by a final fully-connected layer ($FC_D$). The overall mapping is:
\begin{align}
x_D=D_{\theta_D}(x),
\end{align}
where $x$ is an actual ($x_r$) or synthetic image ($\hat{x}$), $x_D\in \mathbb{R}^{1}$ is the output, and $D_{\theta_D}$ is parametrized by $\theta_D$.

\subsubsection{Training of the Global MRI Prior}
To learn a global MRI prior, FedGIMP performs federated training of the unconditional adversarial model (Alg. \ref{alg:flgimp_training}). The training lasts a total of $L$ communication rounds between the sites and the server. A shared generator ($G$) and $K$ local generator copies ($G^k$) are maintained. Meanwhile, $K$ local discriminators ($D^k$) are not exchanged to limit communication load and augment data privacy. In the first round, $G$ and $D^k$ are randomly initialized. At the start of each round, local generators are initialized with the global generator broadcast from the server, $\theta_G^k \gets \theta_G$. In each round, $I$ local epochs are performed to update local models. The local generator updates are calculated to minimize a non-saturated logistic adversarial loss ($L^{k}_{G}$):
\begin{equation}
\label{eq:fedgimp_synthesizer_loss}
L^{k}_{G}(\theta^{k}_G) = -\mathbb{E}_{p(z)}\left[\log(f(D^{k}(G^{k}_{\theta^{k}_G}(z\oplus v,n))\right],
\end{equation}
where $\mathbb{E}_{p(.)}$ is expectation with respect to probability distribution $p$. The local discriminator updates are calculated to also minimize a non-saturated logistic adversarial loss ($L^{k}_{D}$), along with a gradient penalty according to the learned distribution of actual MR images $p(x^k_r)$ \cite{StyleGAN1}:
\begin{align}
L^{k}_{D}(\mathcal{D}^k;\theta^{k}_D) = & -\mathbb{E}_{p(z)}\left[ \log(1 - f(D^{k}_{\theta^{k}_D}(G(z\oplus v,n)))\right]\notag \\ 
& - \mathbb{E}_{p(x^k_r)}\left[\log(f(D^{k}_{\theta^{k}_D}(x^k_r))\right] \notag \\  
& + \frac{\delta}{2} \mathbb{E}_{p(x^k_r)}\left[\left\|\nabla D^{k}_{\theta^{k}_D}(x^k_r)\right\|^{2}\right], \label{eq:fedgimp_discriminator_loss}
\end{align}
where $\mathcal{D}^k$ are training data from site $k$, i.e. coil-combined MR images derived from fully-sampled acquisitions ($x^k_r$). After $I$ iterations of updates according to Eqs. \ref{eq:fedgimp_synthesizer_loss} and \ref{eq:fedgimp_discriminator_loss}, local generators are sent to the server for aggregation \cite{McMahan2017CommunicationEfficientLO}:
\begin{align}
\label{eq:fedgimp_avg}
\theta_G =  \sum_{k=1}^{K} \alpha^k \theta_G^k 
\end{align}

\begin{algorithm}[t]\small
\caption{FedGIMP inference}\label{alg:cap}

\KwInput { $y_{test}^{k,s}$: Undersampled data for $k^{th}$ site, $s^{th}$ subject.\\
$A_{test}^{k,s}$: subject-specific imaging operator for $y_{test}^{k,s}$.\\ 
$G=\{ M, S\}$: global generator with params. $\theta^*_G=\{ \theta^*_M, \theta^*_S\}$.\\
$v$: site-specific one-hot encoding vector.\\
$z$: i.i.d normal variable.\\
$n$: randomly initialized noise variable.\\
$E$: number of iterations for inference.\\
}
\KwOutput {$\hat{m}_{k,s}$: reconstructed image.}
$\theta_S^{1} \gets \theta^*_S$, $\theta_M^{1} \gets \theta^*_M$, $n^{1} \gets n$  (initialize)\\
\For{$e=1:E$}
{
            Calculate $\nabla_{\theta_S^{e},\theta_M^{e},n^e}\mathcal{L}^{k,s}_{DC}(y^{k,s}_{test},A_{test}^{k,s};\theta_S^{e},\theta_M^{e},n^e)$ based on Eq. \ref{eq:fedgimp_optimization} \\
            $\theta^{e+1}_{S} \gets \theta^{e}_{S} - Opt(\nabla_{\theta_S^{e}}\mathcal{L}^{k,s}_{DC})$ \\
            $\theta^{e+1}_{M} \gets w^{e}-Opt(\nabla_{\theta^{e}_{M}}\mathcal{L}^{k,s}_{DC})$ \\
            $n^{e+1} \gets n^{e}-Opt(\nabla_{n^e}\mathcal{L}^{k,s}_{DC})$ \\   

}
$\hat{m}_{k,s}=S_{\theta^{E}_{S}}(M_{\theta_M^E}(z \oplus v),n^{E})$\\
\Return $\hat{m}_{k,s}$

\end{algorithm}

\subsubsection{Inference at a Test Site} No information regarding the imaging operators are involved in the calculation of the training losses in Eqs. \ref{eq:fedgimp_synthesizer_loss} and \ref{eq:fedgimp_discriminator_loss}. As such, the trained generator synthesizes high-quality, coil-combined MR images, but it cannot directly map undersampled data to an image. Therefore, to enable reconstruction based on the learned MRI prior, FedGIMP combines it with the subject-specific imaging operator at a test site ($A^{k,s}_{test}$). Note that this combination can be performed by including the prior in the optimization problem for various model-based reconstruction methods \cite{Uecker2014,ADMM-CSNET,raki,MoDl}. Here, we follow a straightforward approach where the prior is adapted to the reconstruction task by minimizing a data-consistency loss (Alg. \ref{alg:cap}):
\begin{align}
L^{k,s}_{DC}(y^{k,s}_{test},A^{k,s}_{test};\theta_S,\theta_M,n)=&\left\|A^{k,s}_{test}S_{\theta_S}(w,n)-y^{k,s}_{test}\right\|_{2} \notag \\ 
 &+ \eta \left\|\nabla S_{\theta_S}(w,n)\right\|, 
\label{eq:fedgimp_optimization}
\end{align} 
where $w = M_{\theta_M}(z \oplus v)$, data-consistency is enforced on acquired k-space samples, and a gradient penalty is included with weight $\eta$ to prevent noise amplification \cite{aggarwal2021}. For each cross-section, the optimization is performed over the synthesizer parameters ($\theta_S$), mapper parameters ($\theta_S$) and noise ($n$). Given the trained global generator $G_{\theta^*_G}$, synthesizer and mapper parameters are initialized as $\theta_S^{1} \gets \theta^*_S$ and $\theta_M^{1} \gets \theta^*_M$, whereas instance-specific noise variables are randomly initialized as $n^{1} \gets n$. Following a total of $E$ iterations, the adapted prior is used to compute the final reconstruction: 
\begin{align}
\hat{m}_{k,s}=S_{\theta^{E}_S}(M_{\theta_M^E}(z \oplus v),n^E)
\label{reconstruction}
\end{align}

\vspace{-2mm}
\section{Methods}
\vspace{-1mm}
\subsection{Architectural Details}
The unconditional adversarial model in FedGIMP used a mapper with 8 FC layers receiving a standard normal vector and a one-hot binary encoding vector of site index as inputs, while outputting 32 intermediate latent variables. A synthesizer with 8 layers was used, where each layer contained a bilinear upsampling block for 2-fold increase of feature map resolution, followed by two serial cascades of Conv, NI and AdaIN blocks. The first layer received a 4x4 map of learnable constant values, initialized with ones, as input. The learnable noise variable was randomly initialized from a standard normal distribution. A discriminator with 8 layers was used, each containing a bilinear downsampling block for 2-fold reduction in resolution, and a convolution block with 3x3 kernel size. Leaky ReLU activation functions were used. Two separate channels at the generator's output and the discriminator's input were used to represent real and imaginary parts of complex MR images. During training, images were zero-padded to match the resolution of the output layer if needed. During inference, synthesized images were centrally cropped based on the size of the acquisition matrix prior to calculation of data-consistency loss. The synthesizer and discriminator were trained non-progressively, with all layers intact. 

\vspace{-4mm}
\subsection{Competing Methods}
\vspace{-1mm}
FedGIMP was demonstrated against a traditional reconstruction (LORAKS), non-federated models (GAN\SB{cond}/GIMP), and federated conditional models (FL-MRCM, FedGAN, LG-Fed, FedMRI). Hyperparameter selection was performed based on validation performance. For each method, a single set of learning rate, number of epochs (non-federated), number of communication rounds and epochs (federated), and loss term weights were selected that yielded near-optimal performance across tasks.  Unconditional models were trained via Adam optimizer at a learning rate of $10^{-3}$, $\beta_1$=0.0, $\beta_2$=0.99, $\delta$=10; conditional models were trained via Adam optimizer at a learning rate of 2x$10^{-4}$, $\beta_1$=0.5, $\beta_2$=0.99. Training lasted 100 epochs for non-federated models, $L$=100 rounds and $I$=1 epochs for federated models. Strict data consistency was enforced on all reconstructions prior to reporting. We implemented LORAKS in Matlab, unconditional models in TensorFlow and conditional models in PyTorch. Models were executed on a system with four Nvidia RTX 3090s.

\underline{\textit{FedGIMP:}} The proposed model was trained to synthesize coil-combined MR images. The model output magnitude images for single-coil experiments, and complex images with real and imaginary components as separate channels for multi-coil experiments. Inference was performed via Adam optimizer at a learning rate of $10^{-2}$, $\eta$=$10^{-4}$ and $E$=1200 iterations.

\underline{\textit{LORAKS:}} A traditional autocalibrated low-rank reconstruction was performed \cite{Haldar2016}. The k-space neighborhood radius and the rank of LORAKS matrix were selected as (2,6) for single-coil data and (2,30) for multi-coil data.

\underline{\textit{GAN\SB{cond}:}} A non-federated conditional model was trained to map zero-filled (ZF) reconstruction of undersampled acquisitions to reference images of fully-sampled acquisitions \cite{rgan}. Weights of (pixel-wise, adversarial, perceptual) losses were set as (100, 1, 100). Centralized training was performed following dataset aggregation across sites, albeit an unshared discriminator was used to process data coming from different sites as in FedGIMP for improved performance. GAN\SB{cond} serves as a privacy-violating benchmark for conditional reconstruction.  

\underline{\textit{GIMP:}} A non-federated unconditional model was trained based on the architecture and loss functions in FedGIMP. Centralized training was performed, other training and inference procedures were identical to FedGIMP. GIMP serves as a privacy-violating benchmark for unconditional reconstruction.  

\underline{\textit{FL-MRCM:}} A federated conditional model was trained with adversarial alignment of latent representations across sites \cite{guo2021}. For fair comparison among FL methods, architecture and loss functions were adopted from GAN\SB{cond}. Latent representations from the residual backbone were passed through a convolution layer and provided to a domain-alignment network. Weights of (pixel-wise, adversarial, perceptual, reconstruction, domain-alignment) losses were set as (100, 1, 100, 0.5, 0.5).

\underline{\textit{FedGAN:}} A federated conditional model was trained with a shared encoder and decoder across sites \cite{FedGAN}. The architecture and loss functions followed GAN\SB{cond}. Weights of (pixel-wise, adversarial, perceptual) losses were set as (100, 1, 100).

\underline{\textit{LG-Fed:}} A federated conditional model was trained with site-specific encoders and a shared decoder \cite{liang2020think}. The architecture and losses followed GAN\SB{cond}. Weights of (pixel-wise, adversarial, perceptual) losses were set as (100, 1, 100). 

\underline{\textit{FedMRI:}} A federated conditional model was trained with a shared encoder and site-specific decoders \cite{feng2021}. The architecture and losses followed GAN\SB{cond}, albeit an added contrastive loss was included \cite{feng2021}. Weights of (pixel-wise, adversarial, perceptual, contrastive) losses were set as (100, 1, 100, 10).


\begin{table}[t]
\vspace{-0.3cm}
\centering
\caption{Reconstruction performance for single-coil datasets with the imaging operator matched across sites, and across the training-test sets. \emph{Boldface indicates the top-performing FL method}.}
\resizebox{0.487\textwidth}{!}
{
\begin{tabular}{| Sc | l l | Sc | Sc | Sc | Sc | Sc | Sc |}
 \hline
& & & \multicolumn{2}{Sc|}{IXI} & \multicolumn{2}{Sc|}{fastMRI}& \multicolumn{2}{Sc|}{BRATS}\\
 \hline
{} & &  &PSNR $\uparrow$   & SSIM$\uparrow$    & PSNR $\uparrow$   & SSIM$\uparrow$    & PSNR $\uparrow$  & SSIM$\uparrow$   \\
 \hline
\multirow{6}{*}{\rotatebox[origin=c]{90}{\textbf{Non-fed}}}& \multirow{2}{*}{LORAKS}   & 3x & 28.58 & 69.88 & 29.08 & 70.40 & 35.10 & 95.08 \\
& & 6x & 25.98 & 61.87 & 25.95 & 62.04 & 31.71 & 92.29\\
 \cline{2-9}
& \multirow{2}{*}{GAN\SB{cond}}   & 3x & 38.62 & 94.58 & 39.19 & 95.11 & 44.08 & 98.46 \\
& & 6x &35.12 & 91.18 & 35.83 & 91.94 & 40.25 & 97.14 \\
 \cline{2-9}
& \multirow{2}{*}{GIMP}   & 3x&42.21 & 98.23 & 40.56 & 97.06 & 50.55 & 99.65 \\
& &6x&37.56 & 96.42 & 36.62 & 94.45 & 45.38 & 99.17 \\
 \hline
\multirow{10}{*}{\rotatebox[origin=c]{90}{\textbf{Fed}}} & \multirow{2}{*}{FL-MRCM}   & 3x&38.10& 94.19 & 38.75 & 94.59 & 43.44 & 98.27 \\
& &6x&34.72 & 90.75 & 35.26 & 90.92 & 39.58 & 96.81 \\
 \cline{2-9}
& \multirow{2}{*}{FedGAN}  &3x& 38.47 & 94.51 & 38.90 & 94.86 & 44.07 & 98.43  \\
&  &  6x& 35.04 & 91.20 & 35.64 & 91.68 & 40.10 & 96.96 \\
 \cline{2-9}
{} & \multirow{2}{*}{LG-Fed}   & 3x& 38.42 & 94.27 & 38.94 & 94.92 & 44.04 & 98.43\\
& & 6x& 34.80 & 90.63 & 35.57 & 91.61 & 40.10 & 97.07\\
  \cline{2-9}
& \multirow{2}{*}{FedMRI}  &3x& 38.61 & 94.44 & 39.00 & 94.91 & 44.23 & 98.49  \\
& & 6x& 34.85 & 90.47 & 35.54 & 91.42 & 40.34 & 97.20\\
 \cline{2-9}

& \multirow{2}{*}{FedGIMP}   & 3x& \textbf{42.89} & \textbf{98.46} & \textbf{41.06} & \textbf{97.36} & \textbf{50.88} & \textbf{99.69}\\
& & 6x& \textbf{37.75} & \textbf{96.52} &	\textbf{36.92} & \textbf{94.76} & \textbf{45.35} & \textbf{99.15}\\
\hline
\end{tabular}
}
\label{tab:within_domain_sc}
\end{table}

\vspace{-0.2cm}
\subsection{Experiments}
\vspace{-0.1cm}
\subsubsection{Datasets}
FL experiments were conducted using an in-house dataset acquired at Bilkent University along with three public datasets (IXI {\small (\url{http://brain-development.org/ixi-dataset/})}, fastMRI \cite{fastmri}, BRATS \cite{BRATS}). We describe the in-house imaging protocol below (see related references for public datasets). Acquisitions were retrospectively undersampled using variable- (VD) and uniform-density (UD) patterns at acceleration rates R= (3x, 6x) \cite{Lustig2007}. There was no subject overlap between training, validation, and test sets. 

\underline{\textit{In-House:}} T\SB{1}-, T\SB{2}- and PD-weighted scans were performed in 10 subjects on a 3T Siemens Tim Trio scanner located at Bilkent University using a 32-channel coil. An MP-RAGE sequence was used for T\SB{1}-weighted scans with TE/TI/TR = 3.87/1100/2000 ms, 20$^o$ flip angle; and an FSE sequence was used for T\SB{2}-/PD-weighted scans
with TE$_{PD}$/TE$_{T2}$/TR = 12/118/1000 ms, 90$^o$ flip angle. All scans were performed with 192x256x176 mm$^3$ field-of-view and 1x1x2 mm$^3$ voxel size. Ethics approval was obtained from the local ethics committee, and all participants gave written-informed consent.

\begin{table}[t]
\vspace{-0.3cm}
\centering
\caption{Reconstruction performance for single-coil datasets with the imaging operator mismatched across the training-test sets, $A \rightarrow B$ denotes the domain shift in acceleration rates (upper panel) / sampling densities (lower panel).}
\resizebox{0.487\textwidth}{!}
{
\begin{tabular}{| Sc | l l |  Sc | Sc | Sc | Sc | Sc | Sc | }
\hline
\multicolumn{9}{|Sc|}{\textbf{Acceleration Rate} (6x$\rightarrow$3x or 3x$\rightarrow$6x)} \\
\hline
 & & & \multicolumn{2}{Sc|}{IXI} & \multicolumn{2}{Sc|}{fastMRI}& \multicolumn{2}{Sc|}{BRATS}\\
\hline
{} & &  &PSNR $\uparrow$   & SSIM$\uparrow$    & PSNR $\uparrow$   & SSIM$\uparrow$    & PSNR $\uparrow$  & SSIM$\uparrow$  \\
 \hline
\multirow{6}{*}{\rotatebox[origin=c]{90}{\textbf{Non-fed}}}& \multirow{2}{*}{LORAKS}   & \textcolor{white}{6x$\rightarrow$}3x & 28.58 & 69.88 & 29.08 & 70.40 & 35.10 & 95.08\\
&& \textcolor{white}{3x$\rightarrow$}6x & 25.98 & 61.87 & 25.95 & 62.04 & 31.71 & 92.29\\
  \cline{2-9}
& \multirow{2}{*}{GAN\SB{cond}}   & 6x$\rightarrow$3x & 37.78 & 92.94 & 38.02 & 93.72 & 42.23 & 97.83\\
&& 3x$\rightarrow$6x & 33.54 & 89.61 & 34.31 & 90.21 & 39.29 & 96.88\\
 \cline{2-9}
& \multirow{2}{*}{GIMP}   & \textcolor{white}{3x$\rightarrow$}3x& 42.21 &	98.23 &	40.56 &	97.06 &	50.55 &	99.65\\
&&\textcolor{white}{3x$\rightarrow$}6x& 37.56 &	96.42 &	36.62 &	94.45 &	45.38 &	99.17\\
 \hline
\multirow{10}{*}{\rotatebox[origin=c]{90}{\textbf{Fed}}}& \multirow{2}{*}{FL-MRCM}  & 6x$\rightarrow$3x& 37.41 & 92.59 & 37.30 & 92.77 & 40.71 & 96.88\\
 &&  3x$\rightarrow$6x& 33.08 & 88.45 & 33.93 & 89.11 & 38.58 & 96.39\\
 \cline{2-9}
 & \multirow{2}{*}{FedGAN}  &6x$\rightarrow$3x& 37.52 &	92.72 &	38.05 &	93.75 &	41.90 &	97.63\\
 &&  3x$\rightarrow$6x& 33.30	& 89.16 &	34.10 &	89.77	& 39.14	& 96.80\\
 \cline{2-9}
&\multirow{2}{*}{LG-Fed}   & 6x$\rightarrow$3x& 37.62 &	92.85 &	37.80 &	93.44 &	42.19 &	97.80\\ 
&& 3x$\rightarrow$6x& 33.92 &	89.87 &	34.33 &	90.16 &	39.57 &	97.03\\
 \cline{2-9}
&\multirow{2}{*}{FedMRI}  & 6x$\rightarrow$3x& 37.64 &	92.79 &	37.79 &	93.48 &	42.68 &	98.00\\
 &&   3x$\rightarrow$6x& 34.08 &	90.15 &	34.29 &	89.91 &	39.61 &	97.05\\
 \cline{2-9}

&\multirow{2}{*}{FedGIMP}   & \textcolor{white}{6x$\rightarrow$}3x& \textbf{42.89} & \textbf{98.46} & \textbf{41.06} & \textbf{97.36} & \textbf{50.88} & \textbf{99.69}\\
&& \textcolor{white}{3x$\rightarrow$}6x& \textbf{37.75} & \textbf{96.52} & \textbf{36.92} & \textbf{94.76} & \textbf{45.35} & \textbf{99.15}\\
\hline
\multicolumn{9}{|Sc|}{\textbf{Sampling Density} (VD$\rightarrow$UD)} \\
\hline
& & & \multicolumn{2}{Sc|}{IXI} & \multicolumn{2}{Sc|}{fastMRI}& \multicolumn{2}{Sc|}{BRATS}\\

 \cline{2-9}
{} & &  &PSNR $\uparrow$   & SSIM$\uparrow$    & PSNR $\uparrow$   & SSIM$\uparrow$    & PSNR $\uparrow$  & SSIM$\uparrow$ \\
 \hline
\multirow{6}{*}{\rotatebox[origin=c]{90}{\textbf{Non-fed}}}&\multirow{2}{*}{LORAKS}   & 3x & 25.52 &	80.92 &	26.08 &	79.92 &	31.54 &	94.50\\
&& 6x & 21.69 &	67.36 &	21.15 &	63.91 &	24.74 &	85.62\\
 \cline{2-9}
&\multirow{2}{*}{GAN\SB{cond}}   & 3x& 30.31 &	89.53 &	30.97 &	88.71 &	36.99 &	96.14\\
&& 6x & 24.18 &	76.96 &	25.11 &	76.53 &	30.40 &	90.04\\
 \cline{2-9}
 & \multirow{2}{*}{GIMP}   & 3x& 35.31 & 95.72 & 35.07 & 93.83 & 47.04 & 99.52\\
&& 6x & 26.76 &	84.01 &	28.06 &	82.33 &	33.42 &	93.55\\
 \hline
 \multirow{10}{*}{\rotatebox[origin=c]{90}{\textbf{Fed}}}&\multirow{2}{*}{FL-MRCM}   & 3x& 30.45 &	89.83 &	30.65 &	87.47 &	36.06 &	95.07\\
&& 6x & 23.90 &	74.75 &	24.45 &	73.39 &	29.41 &	86.81\\
  \cline{2-9}
 &\multirow{2}{*}{FedGAN}  & 3x& 30.22 &	89.32 &	31.02 &	88.95 &	37.09 &	96.23\\
&& 6x & 24.08 &	76.64 &	25.02 &	76.43 &	30.24 &	89.75\\
 \cline{2-9}
&\multirow{2}{*}{LG-Fed}   & 3x& 30.58 &	89.98 &	30.97 &	88.71 &	36.94 &	96.07\\
&& 6x & 24.24 &	77.40 &	25.08 &	76.45 &	30.19 &	89.85\\
 \cline{2-9}
&\multirow{2}{*}{FedMRI}  & 3x& 30.74 &	90.36 &	31.05 &	88.96 &	37.17 &	96.34\\
&& 6x & 24.32 &	77.82 &	24.97 &	76.05 &	30.43 &	90.30\\
 \cline{2-9}

&\multirow{2}{*}{FedGIMP}   & 3x& \textbf{35.12} & \textbf{95.43} & \textbf{35.33} & \textbf{94.22} & \textbf{47.36} & \textbf{99.55}\\
&& 6x& \textbf{26.59} & \textbf{83.87} & \textbf{27.97} & \textbf{82.28} & \textbf{33.40} & \textbf{93.41}\\
\hline
\end{tabular}
}
\label{tab:cross_domain_sc}
\end{table}

\subsubsection{Single-Coil Reconstruction} Experiments were performed on IXI, fastMRI, and BRATS. For each dataset, multi-contrast acquisitions of (40,10,5) subjects were reserved as (training, validation, test) sets, with 21 cross-sections per contrast randomly selected in each subject. T\SB{1}-, T\SB{2}-, PD-weighted acquisitions in IXI and BRATS, T\SB{1}c-, T\SB{2}-, FLAIR-weighted acquisitions in fastMRI were considered. A total of 2520 cross-sections were used for training.

\subsubsection{Multi-Coil Reconstruction} Experiments were performed on fastMRI brain, fastMRI knee, and In-House brain datasets. For fastMRI brain, T\SB{1}-, T\SB{2}-, FLAIR-weighted acquisitions from (36,6,18) subjects were reserved as (training, validation, test) sets, with 8 cross-sections per contrast randomly selected in each subject. For fastMRI knee, PD-, PDFS-weighted acquisitions from (48,7,24) subjects were reserved, with 9 cross-sections per contrast. For In-House, T\SB{1}-, T\SB{2}-, PD-weighted acquisitions from (6,1,3) subjects were reserved, with 48 cross-sections per contrast. A total of 2592 cross-sections were used for training. Coil compression was performed onto 5 virtual coils \cite{Zhang2013}. Conditional models mapped multi-coil, complex ZF reconstructions of undersampled data to reference images derived from fully-sampled data. GIMP and FedGIMP synthesized complex coil-combined MR images during training \cite{Uecker2014}. The imaging operator was only injected during inference to enforce data consistency based on multi-coil k-space data. 

\vspace{-3mm}
\subsection{Quantitative Assessments}
\vspace{-1mm}
Reconstructed images were compared against reference images obtained via Fourier reconstruction of fully-sampled acquisitions. Peak signal-to-noise ratio (PSNR,dB) and structural similarity index (SSIM,\%) were measured, following normalization of images to [0 1]. Tables list average metrics across test subjects. Statistical significance of differences between methods was assessed via Wilcoxon sign-rank tests.

\begin{figure*}[tp]
\vspace{-0.3cm}
\begin{minipage}{0.2\textwidth}
\caption{Representative reconstructions of a T\SB{2}-weighted acquisition in IXI, a T\SB{1}c-weighted acquisition in fastMRI, and a T\SB{1}-weighted acquisition in BRATS datasets at R=3x based on uniform-density patterns. Results are shown for ZF, LORAKS, FL-MRCM, FedGAN, LG-Fed, FedMRI, and FedGIMP along with the reference images. Conditional models were trained on variable-density patterns at R=3x.}
\captionsetup{justification   = justified,singlelinecheck = false}
\label{fig:singlecoil}
\end{minipage}
\begin{minipage}{0.8\textwidth}
\centerline{\includegraphics[width=0.975\textwidth]{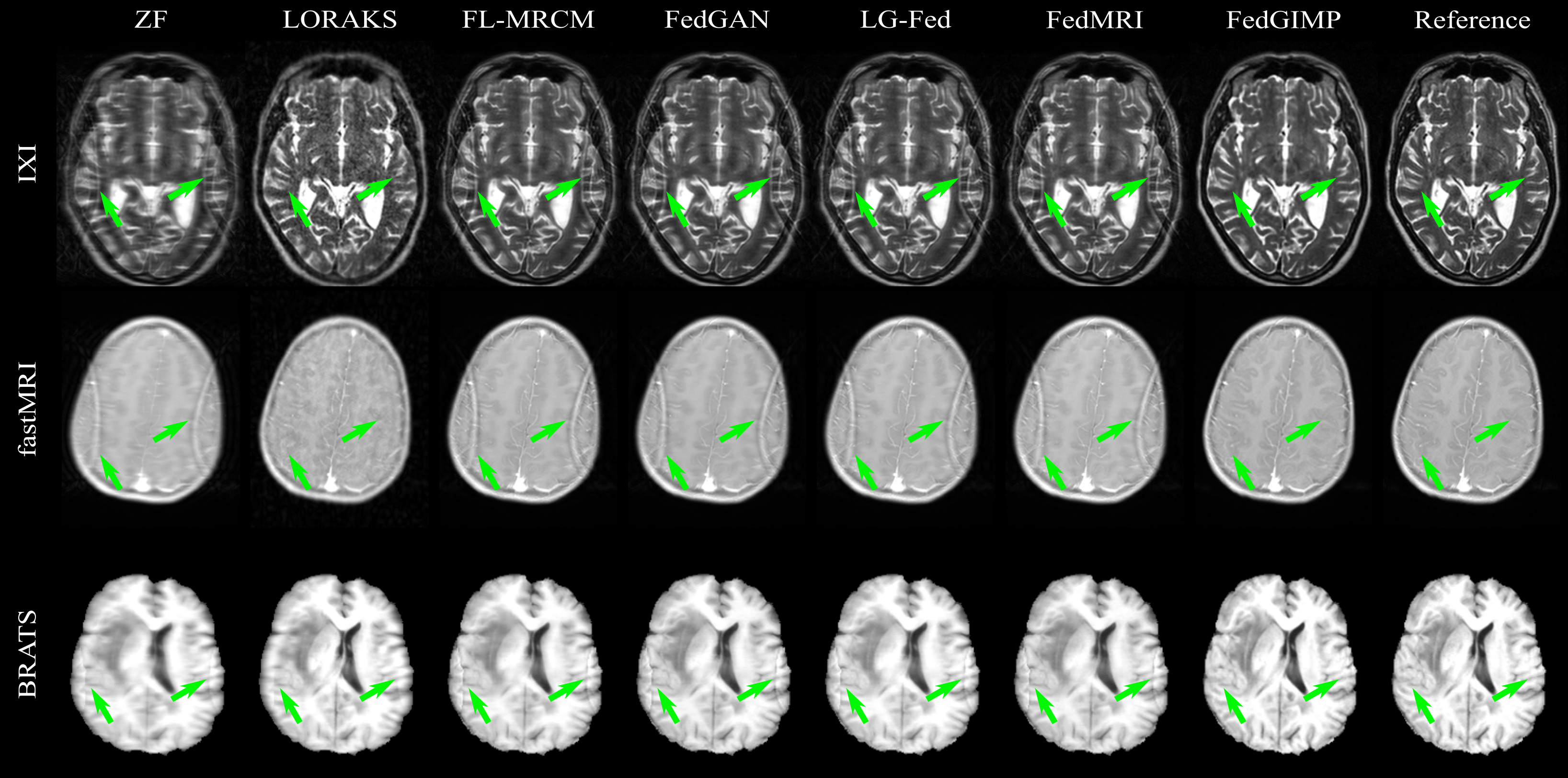}}
\end{minipage}\hfill
\end{figure*}

\vspace{-0.2cm}
\section{Results}
\subsection{Single-Coil Reconstruction}

\subsubsection*{Domain shifts in the MR image distribution}
Multi-site datasets of multi-contrast MRI data contain intrinsic domain shifts in the image distribution within and across sites \cite{guo2021}. To assess the influence of these intrinsic shifts, we examined performance when the imaging operator was matched across sites and across the training-test sets (Table \ref{tab:within_domain_sc}). FedGIMP outperforms all competing methods (p$<$0.05), except on BRATS, R=6x where the privacy-violating benchmark GIMP performs similarly. On average across sites and R, FedGIMP yields 3.66 dB PSNR, 3.00\% SSIM improvement over the second-best FL method, demonstrating the efficacy of FedGIMP against heterogeneity in the MR image distribution.     

\subsubsection*{Domain shifts in the imaging operator}
We then examined reconstruction performance under additional domain shifts due to the imaging operator. First, we considered homogeneous imaging operators across sites, while either acceleration rates (e.g., training at 3x, testing at 6x) or sampling densities (training with VD, testing with UD patterns) were mismatched across the training and test tests (Table \ref{tab:cross_domain_sc}). FedGIMP is the top performer among all methods (p$<$0.05), except for GIMP that performs similarly. Compared to the second-best FL method, FedGIMP offers 4.75 dB PSNR, 4.00\% SSIM improvement under mismatched R, and 4.50 dB PSNR, 4.76\% SSIM improvement under mismatched sampling density. Representative reconstructions under mismatched sampling density are shown in Fig. \ref{fig:singlecoil}. Federated conditional models and LORAKS yield prominent aliasing artifacts and blurring, whereas FedGIMP achieves high visual acuity and improved artifact suppression due to its enhanced generalization performance.  

Next, we considered heterogeneous imaging operators across sites for improved flexibility in collaborations. We separately examined performance when the heterogeneous operators were matched or mismatched between the training-test sets (Table \ref{tab:variable_A}). When the training-test operators match, FedGIMP achieves the highest performance among competing methods (p$<$0.05), except for GIMP that performs similarly, and in BRATS where LG-Fed performs slightly better and GAN\SB{cond}, FedGAN, FedMRI perform similarly. When the training-test operators mismatch, FedGIMP achieves the highest performance among competing methods (p$<$0.05), except GIMP that performs similarly. Compared to the second-best FL method, FedGIMP offers 1.31 dB PSNR and 2.70\% SSIM improvement under matched operators, and 5.08 dB PSNR, 2.98\% SSIM improvement under mismatched operators.  

\begin{table}[]
\centering
\caption{Reconstruction performance for single-coil datasets with heterogeneous imaging operators across sites. The operators were either matched (upper panel) or mismatched (lower panel) across the training-test sets.}
\resizebox{0.487\textwidth}{!}
{
\begin{tabular}{| Sc|l  | Sc | Sc | Sc | Sc | Sc | Sc |  }
\hline
\multicolumn{8}{|Sc|}{\textbf{Matched across training-test sets}} \\
\hline
& &  \multicolumn{2}{Sc|}{IXI} & \multicolumn{2}{Sc|}{fastMRI}& \multicolumn{2}{Sc|}{BRATS}\\
& &  \multicolumn{2}{Sc|}{(9x, VD)} & \multicolumn{2}{Sc|}{(3x, VD)}& \multicolumn{2}{Sc|}{(6x, UD)}\\
\hline
{}  & & PSNR $\uparrow$   & SSIM$\uparrow$    & PSNR $\uparrow$   & SSIM$\uparrow$    & PSNR $\uparrow$  & SSIM$\uparrow$   \\
\hline
\multirow{3}{*}{\rotatebox[origin=c]{90}{\textbf{Non-fed}}}&\multirow{1}{*}{LORAKS}    & 25.00 & 58.83 & 29.08 & 70.4 & 24.74 & 85.62\\
 \cline{2-8}
&\multirow{1}{*}{GAN\SB{cond}}    & 33.35 & 88.82 & 38.97 & 94.92 & 33.01 & 93.23\\
 \cline{2-8}
&\multirow{1}{*}{GIMP}   & 35.50 & 94.99 & 40.56 & 97.06 & 33.42 & 93.55\\
\hline
\multirow{5}{*}{\rotatebox[origin=c]{90}{\textbf{Fed}}}&\multirow{1}{*}{FL-MRCM}    & 32.92 & 88.54 & 37.56 & 93.06 & 30.52 & 89.85\\
 \cline{2-8}
&\multirow{1}{*}{FedGAN}  & 33.00 & 88.39 & 38.01 & 93.47 & 32.51 & 92.91\\
 \cline{2-8}
&\multirow{1}{*}{LG-Fed}  &33.35 & 88.79 & 38.92 & 94.86 & \textbf{33.86} & \textbf{94.20}\\
 \cline{2-8}
&\multirow{1}{*}{FedMRI}  & 33.41 & 88.74 & 38.85 & 94.76 & 32.97 & 93.21\\
 \cline{2-8}

&\multirow{1}{*}{FedGIMP}  & \textbf{35.61} & \textbf{95.18} & \textbf{41.06} & \textbf{97.36} & 33.40 & 93.41 \\
\hline
\multicolumn{8}{|Sc|}{\textbf{Mismatched across training-test sets}} \\
\hline
&&   \multicolumn{2}{Sc|}{IXI} & \multicolumn{2}{Sc|}{fastMRI}& \multicolumn{2}{Sc|}{BRATS}\\
&&   \multicolumn{2}{Sc|}{(9x, VD)$\rightarrow$(3x, UD)} & \multicolumn{2}{Sc|}{(3x, VD)$\rightarrow$(6x, UD)}& \multicolumn{2}{Sc|}{(6x, UD)$\rightarrow$(3x, VD)}\\
\hline
{}    &&PSNR $\uparrow$   & SSIM$\uparrow$    & PSNR $\uparrow$   & SSIM$\uparrow$    & PSNR $\uparrow$  & SSIM$\uparrow$   \\
\hline
\multirow{3}{*}{\rotatebox[origin=c]{90}{\textbf{Non-fed}}}&\multirow{1}{*}{LORAKS}    & 25.52 & 80.92 & 21.15 & 63.91 & 35.10 & 95.08\\
 \cline{2-8}
&\multirow{1}{*}{GAN\SB{cond}}   &31.32 & 91.23 & 26.11 & 79.37 & 42.06 & 97.69 \\
 \cline{2-8}
&\multirow{1}{*}{GIMP}   &  35.31 & 95.72 & 28.06 & 82.33 & 50.55 & 99.65\\
\hline
\multirow{5}{*}{\rotatebox[origin=c]{90}{\textbf{Fed}}}&\multirow{1}{*}{FL-MRCM}   &  31.85 & 91.60 & 25.77 & 77.70 & 40.78 & 96.98\\
 \cline{2-8}
&\multirow{1}{*}{FedGAN}  & 32.31 & 92.51 & 25.86 & 79.27 & 40.55 & 96.67\\
 \cline{2-8}
&\multirow{1}{*}{LG-Fed}   & 31.48 & 91.05 & 24.92 & 75.76 & 37.98 & 94.55\\
 \cline{2-8}
&\multirow{1}{*}{FedMRI}  & 31.65 & 91.40 & 24.87 & 75.82 & 37.79 & 94.24\\
 \cline{2-8}
&\multirow{1}{*}{FedGIMP}   &  \textbf{35.12} & \textbf{95.43} & \textbf{27.97} & \textbf{82.28} & \textbf{50.88} & \textbf{99.69}\\
\hline
\end{tabular}
}
\label{tab:variable_A}
\end{table}

\begin{figure*}[tp]
\vspace{-0.3cm}
\begin{minipage}{0.2\textwidth}
\caption{Representative reconstructions of a FLAIR acquisition of the brain in fastMRI, a fat suppressed PD-weighted acquisition of the knee in fastMRI, and a T\SB{1}-weighted acquisition of the brain in In-House datasets at R=6x. Results are shown for ZF, LORAKS, FL-MRCM, FedGAN, LG-Fed, FedMRI and FedGIMP along with the reference images. Conditional models were trained at R=3x.}
\label{fig:multicoil}
\captionsetup{justification   = justified,singlelinecheck = false}
\end{minipage}
\begin{minipage}{0.8\textwidth}
\centerline{\includegraphics[width=0.975\textwidth]{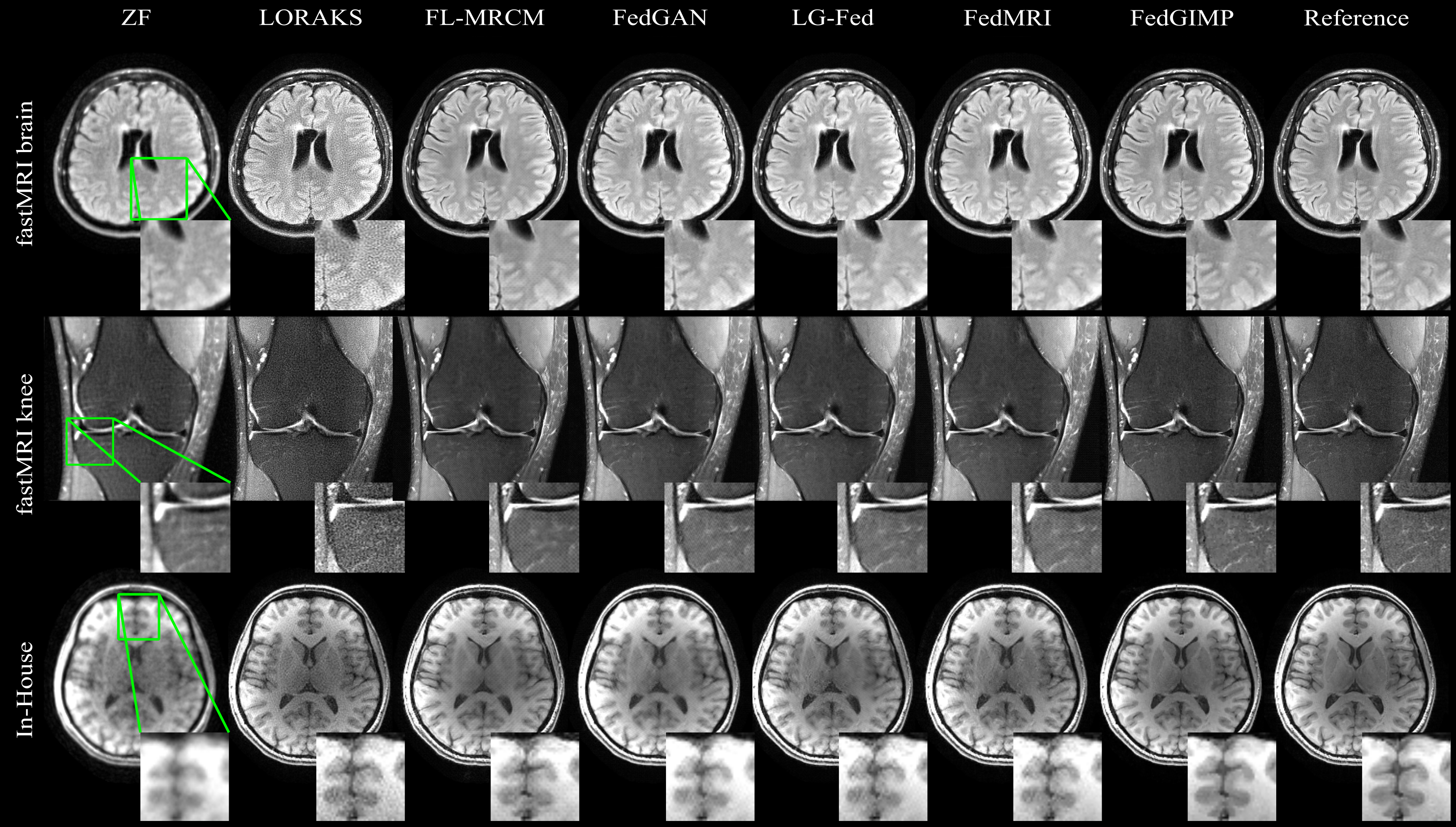}}
\end{minipage}\hfill
\end{figure*}

\begin{table}[h]
\centering
\caption{Reconstruction performance for multi-coil datasets with the imaging operator matched across sites, and the training-test sets.}
\resizebox{0.487\textwidth}{!}
{
\begin{tabular}{| Sc| l l |  Sc | Sc | Sc | Sc | Sc | Sc |  }
\hline
 & & & \multicolumn{2}{Sc|}{fastMRI brain} & \multicolumn{2}{Sc|}{fastMRI knee} & \multicolumn{2}{Sc|}{In-House brain}\\
\hline
{}  & & &PSNR $\uparrow$   & SSIM$\uparrow$ &PSNR $\uparrow$   & SSIM$\uparrow$   & PSNR $\uparrow$   & SSIM$\uparrow$   \\
 \hline
\multirow{6}{*}{\rotatebox[origin=c]{90}{\textbf{Non-fed}}}&\multirow{2}{*}{LORAKS}   & 3x & 36.07 & 89.55 & 32.99 & 78.99 & 51.10 & 99.30\\
& & 6x & 30.81 & 80.29 & 30.88 & 74.09 & 39.28 & 95.67\\
  \cline{2-9}
&\multirow{2}{*}{GAN\SB{cond}}   & 3x & 40.07 &95.56 & 36.08 & 89.85 & 43.46 & 98.66\\
& & 6x & 36.26 & 91.9 & 33.17 & 83.10 & 38.82 & 96.89\\
 \cline{2-9}
&\multirow{2}{*}{GIMP}   & 3x& 47.52 & 98.75 & 43.58&97.49 & 52.43 & 99.62 \\
&& 6x & 40.76 &95.67& 38.02  &91.62& 43.85  &98.50\\
\hline
\multirow{10}{*}{\rotatebox[origin=c]{90}{\textbf{Fed}}}&\multirow{2}{*}{FL-MRCM}   & 3x & 39.40 &95.78 & 36.90 & 90.64 & 40.42 & 97.50\\
& &6x & 35.91 & 92.17 & 33.77 & 82.12 & 35.99 & 94.96\\
 \cline{2-9}
&\multirow{2}{*}{FedGAN}  &3x& 39.83 & 95.60 & 36.95 & 89.96 & 41.70 & 98.11 \\
 &&  6x& 35.92 & 92.06  & 34.01 & 82.70  & 36.66 & 95.67\\
 \cline{2-9}
&\multirow{2}{*}{LG-Fed}   & 3x& 39.76 & 95.74 & 36.56 & 89.54 & 42.76 & 98.48\\
&& 6x& 35.94 & 92.44 & 33.71 & 82.19 & 38.41 & 96.64\\
  \cline{2-9}
& \multirow{2}{*}{FedMRI}  &3x& 39.91 & 95.83 & 36.80 & 89.67 & 43.03 & 98.60 \\
 &&  6x& 36.09 & 92.40 & 33.74 & 81.70 & 38.04 & 96.61\\
 \cline{2-9}

&\multirow{2}{*}{FedGIMP}   & 3x& \textbf{47.23} &\textbf{98.71} & \textbf{43.33} &\textbf{97.38} & \textbf{52.82} &\textbf{99.64}\\
&& 6x& \textbf{40.74} &\textbf{95.71}& \textbf{37.91} &\textbf{91.49} & \textbf{43.96} &\textbf{98.49} \\
\hline
\end{tabular}
}
\label{tab:within_mc}
\end{table}

\vspace{-0.2cm}
\subsection{Multi-Coil Reconstruction}

\subsubsection*{Domain shifts in the MR image distribution}
We first evaluated reconstruction performance when the imaging operator was matched across sites and training-test sets to assess intrinsic domain shifts due to the image distribution. Differing from single-coil experiments, multi-coil FL experiments were conducted on datasets containing two separate anatomies, i.e. the knee and the brain (Table \ref{tab:within_mc}). FedGIMP outperforms all competing methods (p$<$0.05), except GIMP that performs similarly. On average, FedGIMP achieves 6.26 dB PSNR, 4.21\% SSIM improvement over the second-best FL method. These results corroborate the efficacy of FedGIMP in multi-coil settings, and its efficiency in coping with heterogeneity in the image distribution due to diverse anatomy.     

\subsubsection*{Domain shifts in the imaging operator}
We also evaluated reconstruction performance under additional domain shifts due to mismatched imaging operators with different acceleration rates across the training-test sets (Table \ref{tab:cross_mc}). FedGIMP is the top performer among competing methods (p$<$0.05), except for GIMP that performs similarly. FedGIMP offers 6.93 dB PSNR, 4.45\% SSIM improvement over the second-best FL method. Representative reconstructions are shown in Fig. \ref{fig:multicoil}. LORAKS suffers from noise amplification, and conditional models yield notable blur and artifacts. In contrast, FedGIMP achieves high visual acuity while mitigating noise amplification.

\vspace{-0.3cm}
\subsection{Ablation Study}
\vspace{-0.1cm}
Lastly, we conducted an ablation study to assess the contributions of federated training of the MRI prior, and inference adaptation and site-specificity of the proposed mapper to FedGIMP's performance. To do this, we compared FedGIMP against an untrained variant where inference adaptation was performed on a randomly initialized generator, a trained variant that used a static mapper with fixed parameters during inference (i.e., only the synthesizer was adapted), and a trained variant that used a site-general mapper (i.e., latent variables were not site-specific). Performance metrics in Table \ref{tab:ablation} demonstrate that FedGIMP outperforms all ablated variants consistently across sites. These results indicate the importance of the high-quality image prior, subject-specific adaptation of the latent distribution, and site-specific latent distribution in FedGIMP to MRI reconstruction performance.  



\vspace{-0.3cm}
\section{Discussion}
\vspace{-0.1cm}
Multi-site imaging data collected under diverse protocols/devices can contain heterogeneity in the image distribution and the imaging operator across sites, and across the training-test sets \cite{li2021fedbn}. Recent studies have proposed latent-space alignment or split-network approaches based on conditional reconstruction models to address across-site heterogeneity \cite{guo2021,feng2021}. Yet, conditional models are susceptible to domain shifts in the imaging operator pertained to undersampled data \cite{korkmaz2021unsupervised}. In contrast, FedGIMP decouples the undersampling characteristics from the prior to improve reliability against heterogeneity in the imaging operator. Experiments on multi-site datasets demonstrate that FedGIMP yields superior performance against federated conditional models under various imaging scenarios with varying acceleration rates, sampling density across sites, and across the training and test sets. Therefore, FedGIMP can improve flexibility in multi-site collaborations by permitting heterogeneous protocols.   

A practical concern for MRI reconstruction is the computational cost of training and inference. Here, we considered adversarial architectures for unconditional and conditional methods. Previous conditional methods additionally compute either a cross-site alignment loss on latent-space representations \cite{guo2021}, or a weighted-contrastive loss across sites based on local encoder weights \cite{feng2021}. Furthermore, while conditional models are retrained for each different configuration of acceleration rate and sampling density, FedGIMP trains a single MRI prior to reconstruct with various different imaging operators. Thus, FedGIMP offers a simpler training procedure against conditional methods. On the other hand, conditional models offer fast inference in a single forward-pass. In contrast, FedGIMP conjoins the MRI prior with the imaging operator via an iterative procedure that elevates computational burden. This procedure can be accelerated by transfer of optimized model weights across neighboring cross-sections \cite{korkmaz2021unsupervised}.  


\begin{table}[t]
\vspace{-0.3cm}
\centering
\caption{Reconstruction performance for multi-coil datasets with the imaging operator mismatched across training-test sets.}
\resizebox{0.487\textwidth}{!}
{
\begin{tabular}{| Sc| l l |  Sc | Sc | Sc | Sc | Sc | Sc |  }
\hline
& && \multicolumn{2}{Sc|}{fastMRI brain}& \multicolumn{2}{Sc|}{fastMRI knee} & \multicolumn{2}{Sc|}{In-House brain}\\

 \hline
{}  &&  &PSNR $\uparrow$   & SSIM$\uparrow$  &PSNR $\uparrow$   & SSIM$\uparrow$  & PSNR $\uparrow$   & SSIM$\uparrow$  \\
 \hline
 
\multirow{6}{*}{\rotatebox[origin=c]{90}{\textbf{Non-Fed}}}&\multirow{2}{*}{LORAKS} & \textcolor{white}{6x$\rightarrow$}3x & 36.07 & 89.55& 32.99 & 78.99& 51.10 & 99.30\\
&& \textcolor{white}{3x$\rightarrow$}6x & 30.81 & 80.29 & 30.88 & 74.09 & 39.28 & 95.67\\
  \cline{2-9}
&\multirow{2}{*}{GAN\SB{cond}}   & 6x$\rightarrow$3x& 38.98 & 95.14 & 35.76 & 89.46 & 41.49 & 98.44\\
&& 3x$\rightarrow$6x& 36.35 & 92.16 & 33.41 & 83.34 & 37.37 & 95.81\\
  \cline{2-9}
& \multirow{2}{*}{GIMP}   & \textcolor{white}{6x$\rightarrow$}3x& 47.52 & 98.75 & 43.58 & 97.49 & 52.18 & 99.65 \\
&&  \textcolor{white}{3x$\rightarrow$}6x& 40.76 & 95.67 & 38.02 & 91.62 & 43.64 & 98.43\\
 \hline
 \multirow{10}{*}{\rotatebox[origin=c]{90}{\textbf{Fed}}}&\multirow{2}{*}{FL-MRCM}   & 6x$\rightarrow$3x& 37.67 & 94.78 & 35.76 & 88.50 & 37.86 & 96.00\\
&& 3x$\rightarrow$6x& 36.27 & 92.96 & 34.24 & 83.93 & 35.91 & 94.68\\
   \cline{2-9}
&\multirow{2}{*}{FedGAN}  & 6x$\rightarrow$3x& 38.44 & 94.97& 36.12 & 88.68 & 39.69 & 97.67\\
&& 3x$\rightarrow$6x& 35.99 & 92.55 & 34.17 & 83.95 & 36.24 & 95.30\\
  \cline{2-9}
&\multirow{2}{*}{LG-Fed}   & 6x$\rightarrow$3x& 38.56 & 95.10 & 35.75 & 88.16 & 41.31 & 98.14\\
&& 3x$\rightarrow$6x& 36.19 & 92.77 & 34.00 & 83.49 & 37.39 & 96.05 \\
  \cline{2-9}
 &\multirow{2}{*}{FedMRI}  & 6x$\rightarrow$3x& 38.86 & 95.12 & 35.88 & 88.04 & 40.83 & 98.23\\
&& 3x$\rightarrow$6x& 36.23 & 92.79 & 34.07 & 83.32 & 37.60 & 95.89\\
  \cline{2-9}

&\multirow{2}{*}{FedGIMP}   &\textcolor{white}{6x$\rightarrow$}3x& \textbf{47.23} & \textbf{98.71}& \textbf{43.33}&\textbf{97.38} & \textbf{52.82}&\textbf{99.64}\\
&&  \textcolor{white}{3x$\rightarrow$}6x& \textbf{40.74} & \textbf{95.71}& \textbf{37.91}& \textbf{91.49}& \textbf{43.96}& \textbf{98.49} \\
\hline
\end{tabular}
}
\label{tab:cross_mc}
\end{table}

FL methods transfer model weights across sites instead of MRI data to lower privacy risks. Still, security concerns can arise from backdoor attacks where an adversary poisons training updates to corrupt models and elicits diagnostically-inaccurate reconstructions \cite{Kaissis2020}. Previously proposed non-adaptive models freezing model weights following training can be more sensitive to model corruption. In contrast, FedGIMP adapts its MRI prior to each test sample, so it can potentially alleviate corruption during inference \cite{aggarwal2021}. Learning-based models can also be vulnerable to inference attacks on models aiming to leak sensitive information about training data \cite{Kaissis2020}. Differential privacy between training and synthetic samples in adversarial models substantially improves for large and diverse training datasets as encountered in FL settings \cite{FengICCV2021}. Furthermore, FedGIMP uses a shared generator without direct access to data and unshared discriminators that are not communicated \cite{Han2020}. Nonetheless, resilience against inference attacks can be improved by adopting differential-privacy procedures \cite{FedDPGAN,ziller2021}. Future studies are warranted for examining the privacy-preserving abilities of FedGIMP. 

FedGIMP trains an MRI prior over the distribution of high-quality MR images, uninformed regarding the reconstruction task. Thus, the federated prior, in principle, can be adapted for other inverse problems such as MRI super-resolution or synthesis. A simple approach would be to train task-specific models using a synthetic dataset generated via the MRI prior. Alternatively, the trained prior can serve as a non-adaptive plug-and-play regularizer in optimization problems \cite{Konukoglu2019}. To adapt the prior during inference, the imaging operator for the reconstruction task would have to be replaced with corresponding physical signal models in target tasks. 

\vspace{-0.2cm}
\section{Conclusion}
\vspace{-0.1cm}
Here, we introduced a novel MRI reconstruction based on federated learning (FL) of a generative MRI prior and inference adaptation via injection of subject-specific imaging operators onto this prior. Benefits over state-of-the-art federated and traditional methods were demonstrated in multi-site MRI datasets. Improved generalization against domain shifts renders FedGIMP a promising candidate for multi-site collaborations in accelerated MRI. FedGIMP might also be used for physics-based reconstruction in other modalities such as CT, PET, or ultrasound by modifying its imaging operator.

\begin{table}[t]
\vspace{-0.3cm}
\centering
\caption{Reconstruction performance of FedGIMP and ablated variants for single-coil datasets at R=3x. The imaging operator was matched across sites, and across training-test sets.}
\resizebox{0.487\textwidth}{!}
{
\begin{tabular}{| l  |  Sc | Sc | Sc | Sc | Sc | Sc |  }
\hline
 &  \multicolumn{2}{Sc|}{IXI} & \multicolumn{2}{Sc|}{fastMRI} & \multicolumn{2}{Sc|}{BRATS}\\
\hline
{}    &PSNR $\uparrow$   & SSIM$\uparrow$    & PSNR $\uparrow$   & SSIM$\uparrow$   & PSNR $\uparrow$   & SSIM$\uparrow$ \\
 \hline

\multirow{1}{*}{Untrained} & 35.18 & 89.70 & 35.49 & 90.49 & 36.69 & 94.03  \\
\hline
\multirow{1}{*}{Static Mapper} & 40.42 & 97.26 & 39.10 & 95.77 & 46.28 & 
99.27  \\
\hline
\multirow{1}{*}{Site-General Mapper}   & 43.13 & 98.56 & 40.67 & 96.91 & 49.37 & 99.65 \\
\hline
\multirow{1}{*}{FedGIMP}   & \textbf{43.55} & \textbf{98.72} & \textbf{40.88} & \textbf{97.06} & \textbf{49.60} & \textbf{99.67} \\
\hline
\end{tabular}}

\label{tab:ablation}
\end{table}

\vspace{-0.3cm}
\bibliographystyle{IEEEtran}
\bibliography{IEEEabrv,Papers}

\end{document}